\documentclass[conference]{IEEEtran}
\IEEEoverridecommandlockouts
\usepackage{cite}
\usepackage{amsmath,amssymb,amsfonts}
\usepackage{amsthm}
\usepackage{algorithm}
\usepackage{algorithmic}
\usepackage{graphicx}
\usepackage{textcomp}
\usepackage{xcolor}
\usepackage{listings}
\usepackage{multirow} 
\usepackage{booktabs} 
\def\BibTeX{{\rm B\kern-.05em{\sc i\kern-.025em b}\kern-.08em
    T\kern-.1667em\lower.7ex\hbox{E}\kern-.125emX}}

\ifCLASSOPTIONcompsoc
\usepackage[caption=false, font=normalsize, labelfont=sf, textfont=sf]{subfig} 
\else
\usepackage[caption=false, font=footnotesize]{subfig}
\usepackage{geometry}
\makeatother

\begin{document}
\title
{AutoMAS: A Generic Multi-Agent System for Algorithm Self-Adaptation in Wireless Networks
%
\author{\IEEEauthorblockN{
Dingli Yuan\IEEEauthorrefmark{1},
Jingchen Peng\IEEEauthorrefmark{1},
Jie Fan\IEEEauthorrefmark{1},
Boxiang Ren\IEEEauthorrefmark{1}, Lu Yang\IEEEauthorrefmark{2}$^\ddag$ and Peng Liu\IEEEauthorrefmark{2}}
\IEEEauthorblockA{\IEEEauthorrefmark{1}Department of Mathematical Sciences, Tsinghua University, Beijing, China}
\IEEEauthorblockA{\IEEEauthorrefmark{2}Wireless Technology Lab, Central Research Institute, 2012 Labs, Huawei Tech. Co. Ltd., China}
  \IEEEauthorblockA{Email: yanglu87@huawei.com}
\thanks{The first two authors contributed equally to this work and $\ddag$ marked the corresponding author. }
 }
}
\maketitle

\begin{abstract}
%
The wireless communication environment has the characteristic of strong dynamics.
Conventional wireless networks operate based on the static rules with predefined algorithms, lacking the self-adaptation ability.
The rapid development of artificial intelligence (AI) provides a possibility for wireless networks to become more intelligent and fully automated.
As such, we plan to integrate the cognitive capability and high intelligence of the emerging AI agents into wireless networks.
In this work, we propose AutoMAS, a generic multi-agent system which can autonomously select the most suitable wireless optimization algorithm according to the dynamic wireless environment. 
Our AutoMAS combines theoretically guaranteed wireless algorithms with agents' perception ability, thereby providing sounder solutions to complex tasks no matter how the environment changes.
%
%
%
%
As an example, we conduct a case study on the classical channel estimation problem, where the mobile user moves in diverse environments with different channel propagation characteristics.
Simulation results demonstrate that our AutoMAS can guarantee the highest accuracy in changing scenarios.
Similarly, our AutoMAS can be generalized to autonomously handle various tasks in 6G wireless networks with high accuracy.
\end{abstract}


\section{Introduction}
The advancement of artificial intelligence (AI) technologies has enabled the technical feasibility of fully automated and intelligent wireless networks \cite{wang2015artificial}.
Current wireless networks are operated based on static rules, such as 3rd generation partnership project (3GPP) standard, and each vendor owning its core algorithms embedded in its equipments.
Note that due to the lack of intelligence, existing wireless algorithms cannot be changed automatically to keep the optimal performance in the dynamic wireless environments \cite{jiang2025large}.
In the following, we will take the  cloud-radio access network (C-RAN) architecture as an example, to describe the necessity and feasibility of the algorithm self-adaptive scheme.
In C-RAN, baseband units (BBUs) are centralized for processing baseband signals, while remote radio heads (RRHs) are distributed separately to provide radio coverage \cite{rost2014cloud}. 
However, due to spatial and temporal variations across different RRH coverage areas, the fixed algorithms pre-deployed in BBUs become inadequate for maintaining consistent optimal performance across diverse scenarios.
Therefore, a new paradigm for automatically choosing the optimal algorithm to deal with various complicated problems in wireless system is necessitated no matter how the wireless environment changes.
%
%
%
%

With the advancements in large language models (LLMs), AI agents powered by LLMs have demonstrated remarkable cognitive and decision-making capabilities \cite{xu2024survey}, showing great potential for addressing wireless communication tasks \cite{tong2024wirelessagent}.
Among these, the multi-agent systems (MAS), where role-specialized agents collaborate through a structured workflow, have become a hot research focus.
This role-playing mechanism enables better performance in diverse real-world tasks \cite{wu2023autogen} and have been adopted in solving wireless communication problems.
For example, Zou \emph{et al.} proposed a preliminary multi-agent reasoning framework with simulation on an energy saving problem \cite{zou2023wireless}.
Mongaillar \emph{et al.} developed three fixed agents for the power scheduling problem \cite{mongaillard2024large}.
%
%
Additionally, the real-time environmental information can be converted into text data through the integrated perception module \cite{tong2024wirelessagent}, further making the intelligent processing of dynamic and complex wireless tasks possible.

%
However, current LLMs still face difficulties in physical-layer modeling, mathematical derivation, and mathematical operations \cite{guo2024large, shao2024deepseekmath,ahn2024large}. 
Yet, solving actual wireless issues requires end-to-end steps from modeling and analytical derivations to numerical computation, which remain challenging for current LLMs.
Thus, it is not reliable to directly use LLM-powered agents to solve complex communication tasks with high performance.
%
For example, existing MAS for wireless systems simply tackle wireless optimization problems through agent dialogues \cite{zou2023wireless, mongaillard2024large}, which may result in infinite loops or malfunctions due to the lack of theoretical guarantees.
The accuracy and reliability of the outputs from multi-agent conversations are uncertain when MAS face complex and dynamic wireless problems.

To this end, we aim to leverage the theoretically
guaranteed wireless algorithms combined with the perception and cognitive ability of AI agents, addressing wireless problems in an intelligent way.
In this paper, we propose AutoMAS, a generic MAS for algorithm self-adaptation to solve tasks in wireless communication systems.
To simulate the task delegation and resolution process in communication systems, we implement AutoMAS in a C-RAN architecture as an example.
Specifically, the C-RAN infrastructure delegates various baseband processing tasks (e.g., resource allocation, channel estimation, etc.) to AutoMAS.
By perceiving environmental conditions and inferring user intents, AutoMAS dynamically selects suitable algorithms to solve the tasks efficiently.
In AutoMAS, we design a supervisor-executor mechanism, where a suitable set of role-specific agents and their workflow are autonomously generated according to the task and the environment.
This mechanism not only enhances the framework’s ability to address diverse tasks, but also enables fully autonomous execution in dynamic environments.
Then, we provide a case study for channel estimation problem, where the user moves in diverse environments with different propagation characteristics.
Simulations show that our AutoMAS obtains the optimal performance in each scenario, enabling the ability for algorithm self-adaptation.

\section{Multi-agent System for autonomous wireless algorithm adaptations}
\label{sec_MAS_model}
\begin{figure*}
    \centering
    \includegraphics[width=1\linewidth]{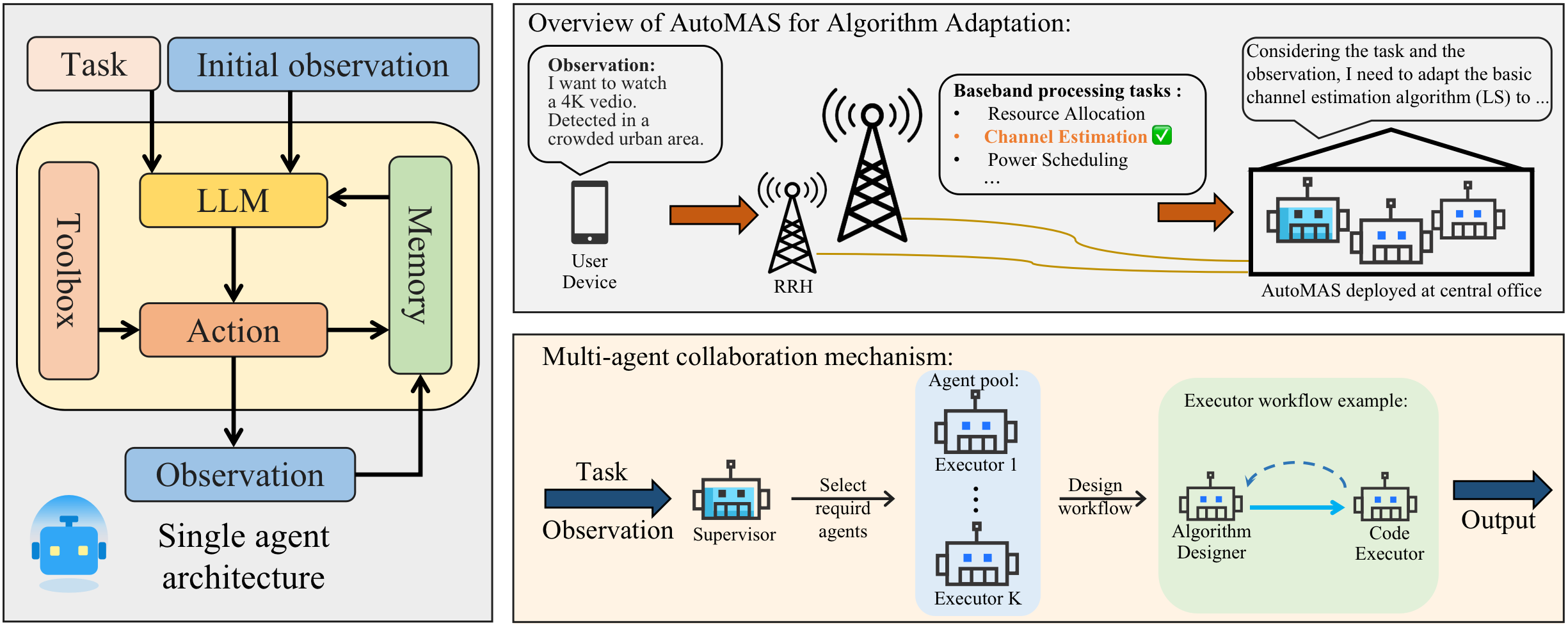}
    \caption{The left part is the single agent architecture. The right part is the overview of AutoMAS for algorithm adaptation and the autonomous multi-agent collaboration mechanism.}
    \label{fig_MAS_model}
\end{figure*}

In this section, we introduce our AutoMAS, a MAS framework for algorithm self-adaptation to handle wireless tasks.
%
%
%
We begin with an overview of AutoMAS. 
Next, we describe its two core components: the single-agent architecture and the multi-agent collaboration mechanism.
Fig. \ref{fig_MAS_model} provides a visual representation that integrates these three key aspects.

\subsection{AutoMAS for RAN-Intelligent Algorithm Adaptation}
\label{subsec_edge_paradigm}
%
%
%
%



Fig. \ref{fig_MAS_model} (top-right) illustrates an overview of AutoMAS, which enables autonomous algorithm adaptation. 
The objective is to enable the wireless networks to have the ability of automatically and intelligently selecting the most suitable algorithms with the environment changing, thereby solving wireless tasks effectively. 
Here, we deploy AutoMAS in a practical C-RAN architecture as an example to validate its operational capabilities.
This architecture centralizes BBUs for baseband processing while distributing RRHs geographically to handle radio frequency (RF) transmission. 
AutoMAS can be deployed at the BBUs and operates centrally, enabling intelligent resolution of baseband processing tasks (e.g., resource allocation, channel estimation, and power control), to maintain consistent performance across heterogeneous and dynamic RRH coverage areas.
The process of AutoMAS consists of two key stages: \textbf{perception} and \textbf{inference}.

In the perception stage, multi-dimensional sensing information about the user's environment within a specific RRH coverage area is observed and collected in textual form. 
This information is denoted as \textbf{observation}, which includes user intent (e.g., a detected request for continuous 4K video playback), environmental information (e.g., radio frequency metrics such as SNR, device states like GPS location and mobility, and obstruction data from radar or cameras), and resource constraints (e.g., available frequency bandwidth, computational resources like GPU memory).
Then, this observation is sent to the AutoMAS for the following processing,  guiding the selection of appropriate algorithms.

In the inference stage, AutoMAS receives a baseband processing task and infers the appropriate algorithmic response based on the observation. 
Specifically, a supervisor agent first analyzes the task and observation, selecting the relevant agents and designing their workflow. 
The selected agents then execute their designated roles and collaborate within this structured workflow. 
Details are further elaborated in next two subsections.
Ultimately, the chosen algorithms are executed, achieving performance-optimized solutions.
%

\subsection{Single Agent Architecture}
\label{subsec_Single_Agent}
Then we introduce the modular design of a single agent.
The agent operates within a closed-loop cognitive architecture, where a LLM coordinates observation, reasoning, and action by tightly integrating memory and external tool.
The core modules are formalized as follows:
\begin{itemize} 
\item \textbf{Task} ($T$) defines the agent’s primary objectives and operational goals.
\item \textbf{Observation} ($O$) serves as the agent’s perception interface with the environment, and consists of: 
\begin{itemize} \item \textbf{Initial observation} ($o_0$) is a multi-dimensional input containing user intent, environmental information and resource constraints. 
\item \textbf{State observation} ($o_t$) captures the system state at step $t$, reflecting both the environmental configuration and the agent's status after an action is executed. 
\end{itemize}
\item \textbf{LLM} ($L$) performs probabilistic reasoning over task specifications, memory contents, and environmental inputs to determine optimal actions.
\item \textbf{Action} ($A$) contains the set of executable instructions the agent can perform to interact with and influence the environment.
It includes default base actions such as \textit{think} for logical reasoning, and \textit{finish} for process termination.
%
Additionally, the action set is expandable to include customized actions, such as \textit{code execution}, which executes the generated code.

\item \textbf{Memory} ($M$) serves as the agent's long-term knowledge repository, encompassing:
\begin{itemize}
\item \textbf{Role definition} specifies operational boundaries (e.g., the \textit{Algorithm Designer} role focuses solely on algorithm development). 
\item \textbf{Domain knowledge} stores expert knowledge for complex tasks, such as a LS estimator. 
\item \textbf{Interaction history} records past action-observation pairs $h_t ={(a_i, o_i)}_{i=1}^{t}$ to support consistent reasoning and lead agent to operate based on the past state and environment information. 
\end{itemize}

\item \textbf{Toolbox} ($T$) provides access to external tools, extending action capabilities. This includes code interpreters (e.g., Python), solvers (e.g., Gurobi), and domain-specific assets such as channel datasets, pre-trained neural network checkpoints and so on.
\end{itemize}
We next present the agent’s internal operational workflow, centered on the \textbf{observation–action chain}:
\begin{enumerate}
    \item Task specifications ($T$) and initial observation ($o_0$) activate the LLM engine.
    \item The LLM selects a probability-maximizing action: $a_t = \arg\max_{a\in A} P(a|T,M,o_0),$ by reasoning over the task ($T$), memory ($M$) and initial observation ($o_0$).
    

    If needed, external tools are invoked from the toolbox during execution.
    \item The execution of $a_t$ modifies the environment, resulting in a new observation $a_t \rightarrow o_t.$

    \item The new action–observation pair is recorded in memory: $M \leftarrow M \cup (a_t,o_t).$
   
    \item The loop continues until termination conditions are met (e.g., executing \textit{finish}).
    \end{enumerate}
This closed-loop interaction allows the agent to take informed action based on both past experiences and current state information, while the integrated modules enhance its ability to handle diverse tasks.

\subsection{Autonomous Multi-agent Collaboration Mechanism}
\label{sub_multi-agent coll}

Existing MAS schemes for tackling with wireless communication tasks are designed for certain scenarios predefined by human \cite{mongaillard2024large}, which thus lacks the self-adaptation ability in dynamic scenarios.
To address this, we introduce an autonomous multi-agent collaboration mechanism implemented in AutoMAS to equip the network with the self-adaptation ability in a wide range of dynamic scenarios.

The effectiveness of AutoMAS relies on two synergistic components: a well-curated \textbf{agent pool} for domain-specific expertise, and a \textbf{supervisor-executor mechanism} for flexible task solving.
The core components are as follows:
\begin{itemize}
\item\textbf{Agent pool}: The agent pool consists of specialized \textbf{executors}, each assigned a distinct role to tackle targeted tasks. 
For example, a communication specialist handles domain-specific analysis, an algorithm designer develops specific algorithms, and a code engineer implements and executes the corresponding solutions. 
These executors are pre-deployed and on standby.
The supervisor dynamically invokes them based on task complexity and environmental observations, ensuring flexibility and efficiency.
%
%
%

\item \textbf{Supervisor–executor mechanism}: In this master–slave architecture, the supervisor acts as the controller, analyzing task semantics and observations to both select executors and orchestrate the workflow.
The selected executors then perform their roles and interact with the others to accomplish the task.
For example, for channel estimation, the supervisor activates two executors: an algorithm selector and a code agent.
The algorithm selector proposes a strategy based on the environment, the code agent implements and validates it, and performance feedback is iteratively fed back to the selector for continuous optimization.
Beyond this simple closed-loop, the supervisor can configure other workflow based on the task characteristics, such as group discussion mode for collective reasoning.
\end{itemize}

By configuring specialized agents' workflows, the supervisor–executor mechanism not only enhances task reasoning efficiency but also achieves automation without rigid predefined patterns, adapting to dynamically changing network environment requirements.

\section{Case Study: Channel Estimation Algorithm Self-adaptation in Dynamic Scenarios }
In this section, we conduct a case study on dynamic channel estimation using AutoMAS to validate its effectiveness.
We first introduce the channel estimation problem and its mathematical formulation.
Then, we describe the primitive configuration of AutoMAS executors for this task.
Next, we present a detailed example illustrating how AutoMAS handles the channel estimation task through agent collaboration.
Finally, we conduct a simulation in diverse scenarios to validate the self-adaptation ability of AutoMAS.

\subsection{Pilot-Aided Channel Estimation Formulation in C-RAN}
Accurate channel state information (CSI) is essential for C-RAN systems, as it directly affects the design of optimal precoders, energy-efficient resource allocation, and other key signal processing tasks\cite{he2018compressive}.
Consider a C-RAN system with $G$ RRHs and a single multi-antenna user device. There are $N_r$ antennas in each RRH and $N_t$ antennas in user device. We introduce the standardized formulation for pilot-aided time-varying channel estimation.
The received signal $Y_{n}^g \in \mathbb{C}^{N_r\times 1}$ at the $g$-th RRH, corresponding to the $n$-th time slot, is given by:
\begin{equation}
Y_{n}^g=H_{n}^g s_{n}+N_{n}^g,   
\end{equation} 
where $s_{n}\in \mathbb{C}^{N_t \times 1}$ denotes the vector of the transmitted pilot, $H_{n}^g \in \mathbb{C}^{N_r\times N_t}$ is the channel matrix between the user and the $g$-th RRH, and $N_{n}^g\in \mathbb{C}^{N_r \times 1}$ is the complex noise vector with zero mean and variance $\sigma^{2}$. 
The objective is to estimate channel matrix $H_{n}^g$ with the known received signal, transmitted pilot and the distribution of the noise.

Existing research on channel estimation has proposed various solutions, ranging from iterative mathematical algorithms to pre-trained AI models \cite{kaur2018channel} \cite{Peaceman}.
However, due to the diversity and variability of wireless environments, fixed methods often fail to adapt effectively \cite{berger2009sparse}.
In the following, we apply AutoMAS for estimation, enabling the autonomous selection of suitable algorithms based on environmental changes to achieve robustness and efficiency.
%

\subsection{Main Executor Configuration Details}
In this subsection, we detail the configuration of the executors within AutoMAS’s agent pool, including their roles, required domain knowledge in memory, and external tools.  

\textbf{Algorithm selector:} This executor chooses the most accurate and efficient estimation algorithm based on user intent, environmental context, and available resources. 
Recognizing current LLM limitations in physical-layer modeling \cite{guo2024large}, we equip this agent’s memory with the core methodologies of pilot-aided channel estimation approaches as basic knowledge. 
Specifically, we roughly categorize these approaches into four groups:
\begin{itemize}
    \item No Prior Assumptions: It requires no statistical knowledge. A typical approach is the LS estimator, which features low computational complexity and wide applicability but exhibits limited noise robustness.
    
    \item Feature-Driven: It utilizes abstract channel features (e.g., sparsity). For instance, compressed sensing methods like Iterative Shrinkage Thresholding Algorithm (ISTA) \cite{beck2009fast} obtain satisfactory accuracy under valid assumptions, but are sensitive to the channel characteristics.
    
    \item Statistics-Driven: It relies on accurate channel statistics (e.g., covariance matrix). For example, Linear Minimum Mean Square Error (LMMSE) \cite{MMSE} provides strong noise resilience, but comes with high computational cost and requires prior knowledge.
    
    \item Data-Driven: It requires a ground-truth dataset. For example, ResNet \cite{he2016deep}, a classical neural network, learns implicit patterns from data, but needs large training datasets and faces generalization challenges.
\end{itemize}
The algorithm selector stores these classification templates in memory and uses them as prompts to guide selection.
%
For example, if the user’s location is identified as an open area, the executor infers a sparse multipath characteristic and selects ISTA to exploit the sparsity.

\textbf{Code agent:} This executor converts algorithm specifications into executable code and runs it using external interpreters. 
%
Its memory stores code snippets that are used as prompts to instantiate and invoke LS, ISTA, LMMSE, or ResNet.
Additionally, its LLM-based reasoning dynamically translates environmental parameters and validation requirements into corresponding code.
For example, when deploying ISTA, it translates descriptors such as “64-antenna configuration” and “20 dB SNR” into code variables (\texttt{antenna\_num = 64}, \texttt{snr\_level = 20}), generates the corresponding sparse channel estimation module, and executes it via the code interpreter.  
%
%
If negative feedback is observed (e.g., poor estimation performance), the code agent packages the relevant channel observations into a diagnostic report and sends it to the algorithm selector to trigger algorithm reselection.
This cross-agent feedback loop, with the goal of enhancing accuracy and efficiency, ensures continual alignment between algorithmic theory and implementation constraints.



\subsection{AutoMAS-Enabled Channel Estimation}
\begin{figure*}
    \centering
    \includegraphics[width=1.0\linewidth]{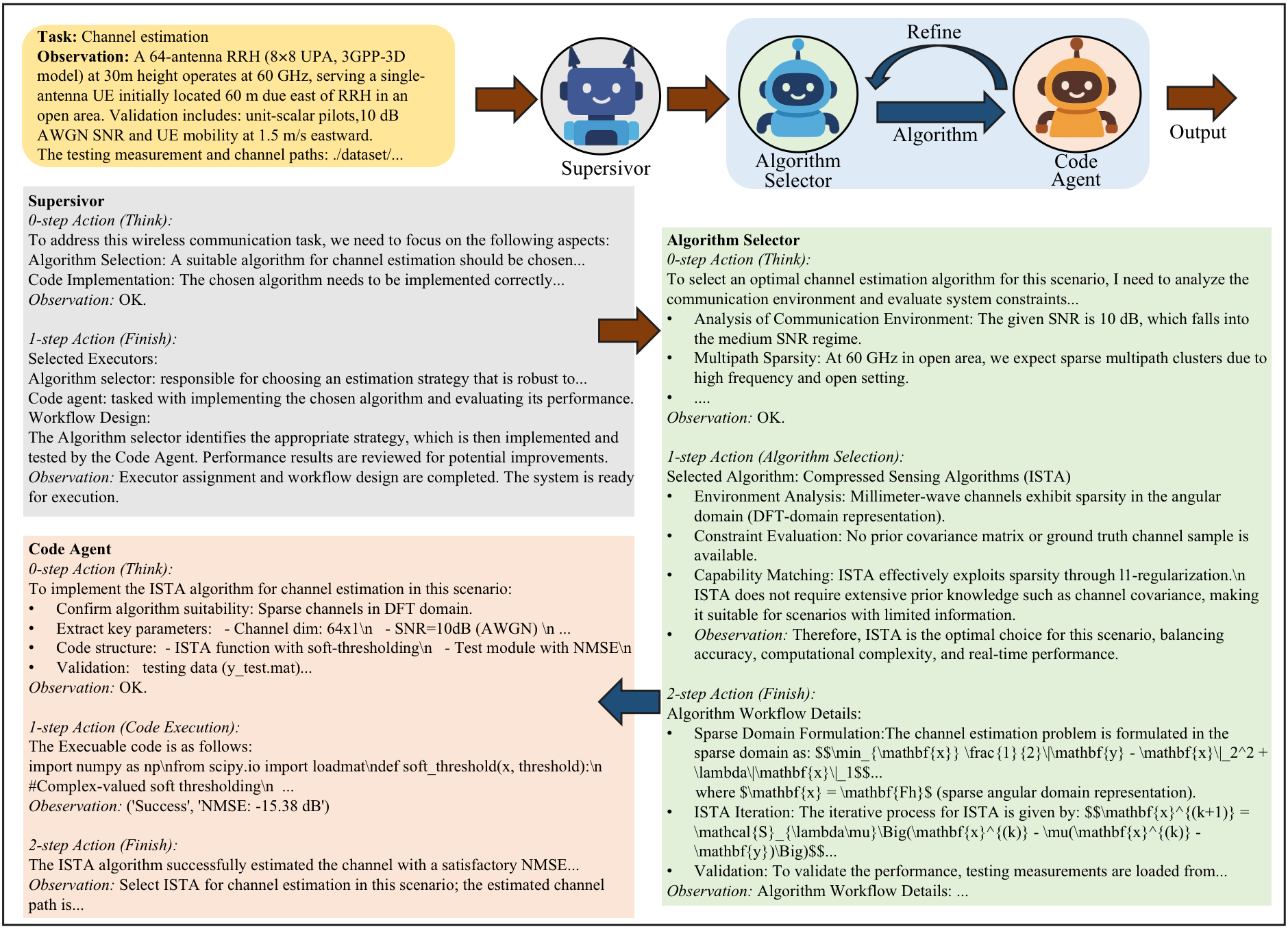}
    \caption{Operation and simplified prompts of AutoMAS for channel estimation}
    \label{fig_example}
\end{figure*}
\label{sec:simulation}
In this subsection, we provide a detailed example to demonstrate AutoMAS’s practical capabilities.
We adopt the QuaDRiGa toolbox \cite{quadriga} as the channel simulator to generate training, validation and testing data.
%
%
%
%
We consider a scenario where the user device (with a single antenna) is located in an open area, served by a 64-antenna RRH installed at a height of 30 meters and positioned 60 meters away, operating at 60 GHz.
The RRH uses a “3GPP-3D” antenna array (AA) with $8\times 8$ elements, 0.5-wavelength spacing, and vertical polarization.
The validation requirements specify physical-layer parameters, including a unit-scalar pilot design, and an AWGN channel with 10 dB SNR, along with the testing channel path for validation. 
These observations, together with the instruction to perform channel estimation, are first sent to the supervisor in AutoMAS for the following processing.
The LLM for simulations uses qwen2.5-vl-72b-instruct.
We use the Normalized Mean Squared Error (NMSE) to measure the performance: $\text{NMSE} = 10\times lg(\|h-\hat{h}\|_2^2/\|h\|_2^2)$, where $h$ is the estimated channel, and $\hat{h}$ is the ground truth channel.
Fig. \ref{fig_example} illustrates the overall process for the scenario and highlights the key prompts associated with each agent, presented in a simplified form for clarity.

The process begins with the supervisor, which receives task specifications and environmental observations. It is responsible for selecting appropriate executors from the agent pool and orchestrating their workflow. 
As illustrated in Fig. \ref{fig_example}, the supervisor activates two executors: the algorithm selector and the code agent, and designs a closed-loop structure to guide their interaction for accomplishing the estimation task. 
%

During the executor interaction phase, the algorithm selector first processes the task based on its specific role. 
It analyzes the environmental conditions and infers a sparse multipath structure, typical for high-frequency (60 GHz) propagation in open terrain. 
Based on this analysis, it selects a compressed sensing method (ISTA) as the estimation strategy. 
%

Next, the code agent implements the selected algorithm with executable code, and invokes an external test module for validation. 
%
As shown in Fig. \ref{fig_example}, the code agent successfully implements the selected algorithm and estimates the channel. 
%
%
The achieved performance is -15.3 dB, indicating satisfactory estimation accuracy without the need for the refinement loop.
%

%




\subsection{Further Comparison in Dynamic Scenarios}
We compare the estimation performance of AutoMAS against baseline methods (LS, ISTA, LMMSE, ResNet) across diverse scenarios to simulate their adaptability to environmental changes induced by user mobility. These scenarios are generated by QuaDRiGa, including dense area with low noise, open area, dense area with high noise, and indoor office .
%
Each scenario is configured with the same “3GPP-3D” AA (8×8 elements, 0.5-wavelength spacing, vertical polarization), but differs significantly in scattering environments, multipath characteristics, carrier frequencies, and other channel parameters.
In addition, we assume that users move at a speed of 1.5 m/s. The specific validation settings for each scenario are summarized as follows.



\begin{itemize}
    \item \textbf{Scenario 1:} Dense area with low noise, 15 GHz, SNR = 20 dB, initial user distance from AA = 50 meters, AA height = 30 meters.
    \item \textbf{Scenario 2:} Open area, 60 GHz, SNR = 10 dB, initial user distance from AA = 60 meters, AA height = 30 meters.
    \item \textbf{Scenario 3:} Dense area with high noise, 15 GHz, SNR = 2 dB, initial user distance from AA = 50 meters, AA height = 30 meters, with only 1,000 ground-truth channel samples available.
    \item \textbf{Scenario 4:} Indoor office, 15 GHz, SNR = 10 dB, initial user distance from AA = 20 meters, AA height = 3 meters, with 30,000 ground-truth samples available.
\end{itemize}

\begin{figure}[t]
\centering
\subfloat[Scenario 1]{\includegraphics[width=0.49\columnwidth]{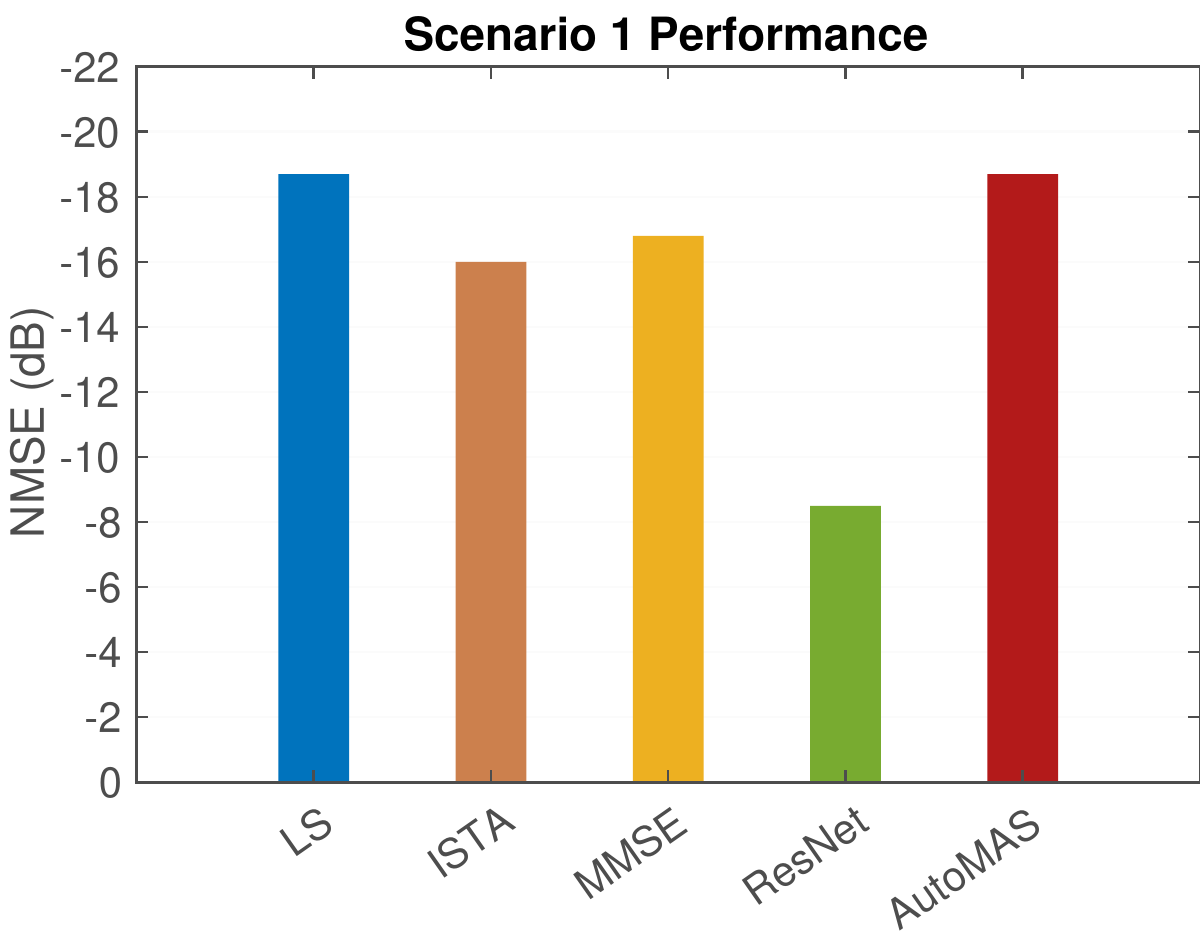}\label{fig:sub1}}
\hfill
\subfloat[Scenario 2]{\includegraphics[width=0.49\columnwidth]{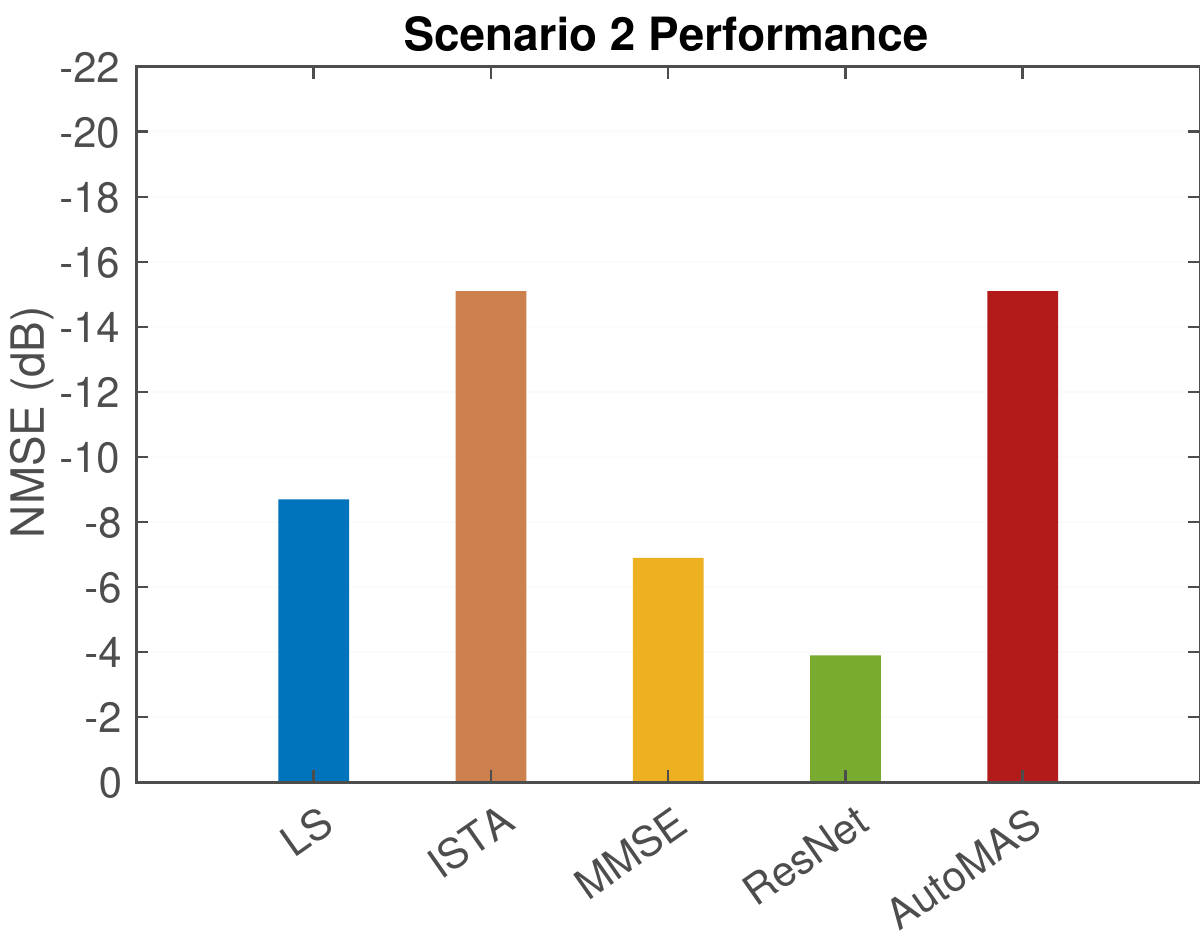}\label{fig:sub2}}

\vspace{0.1cm} 

\subfloat[Scenario 3]{\includegraphics[width=0.49\columnwidth]{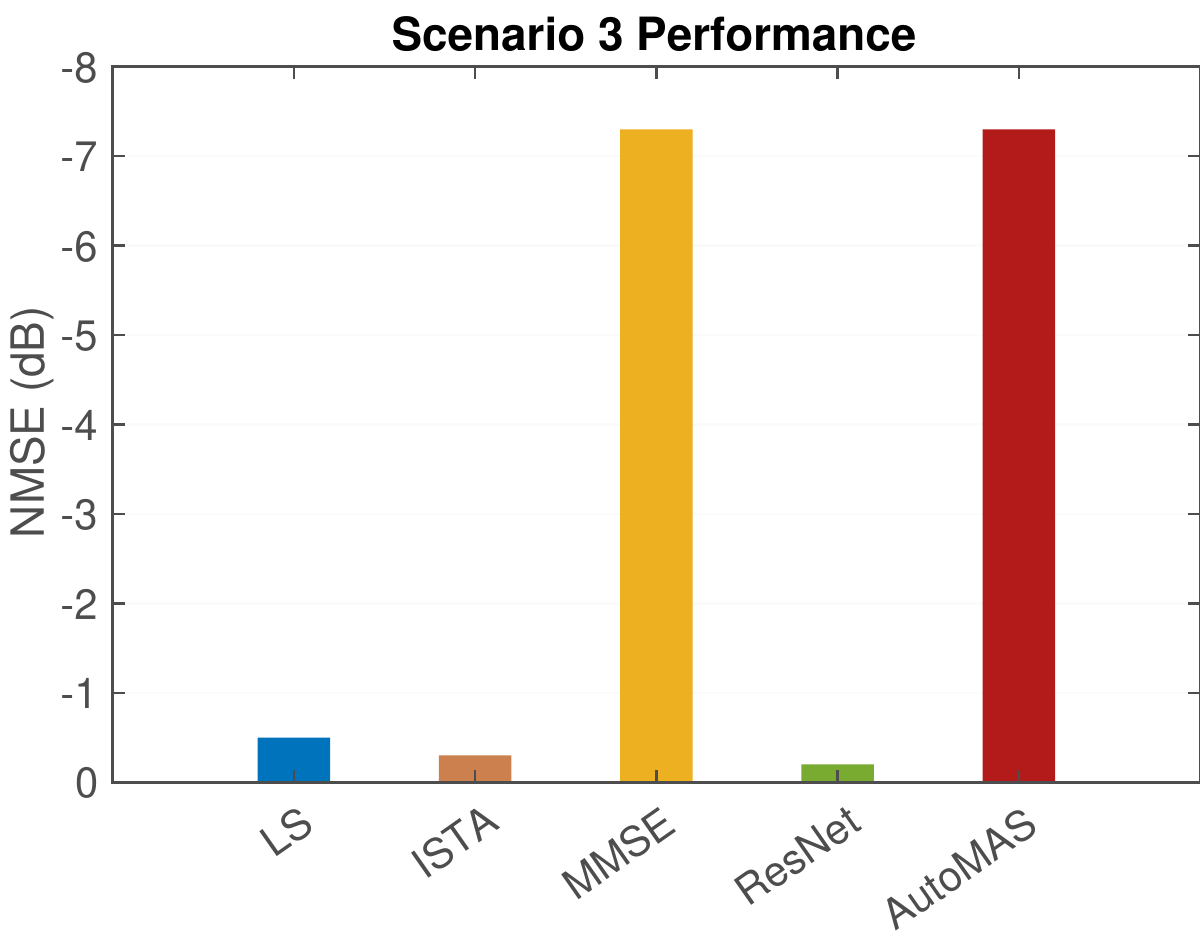}\label{fig:sub3}}
\hfill
\subfloat[Scenario 4]{\includegraphics[width=0.49\columnwidth]{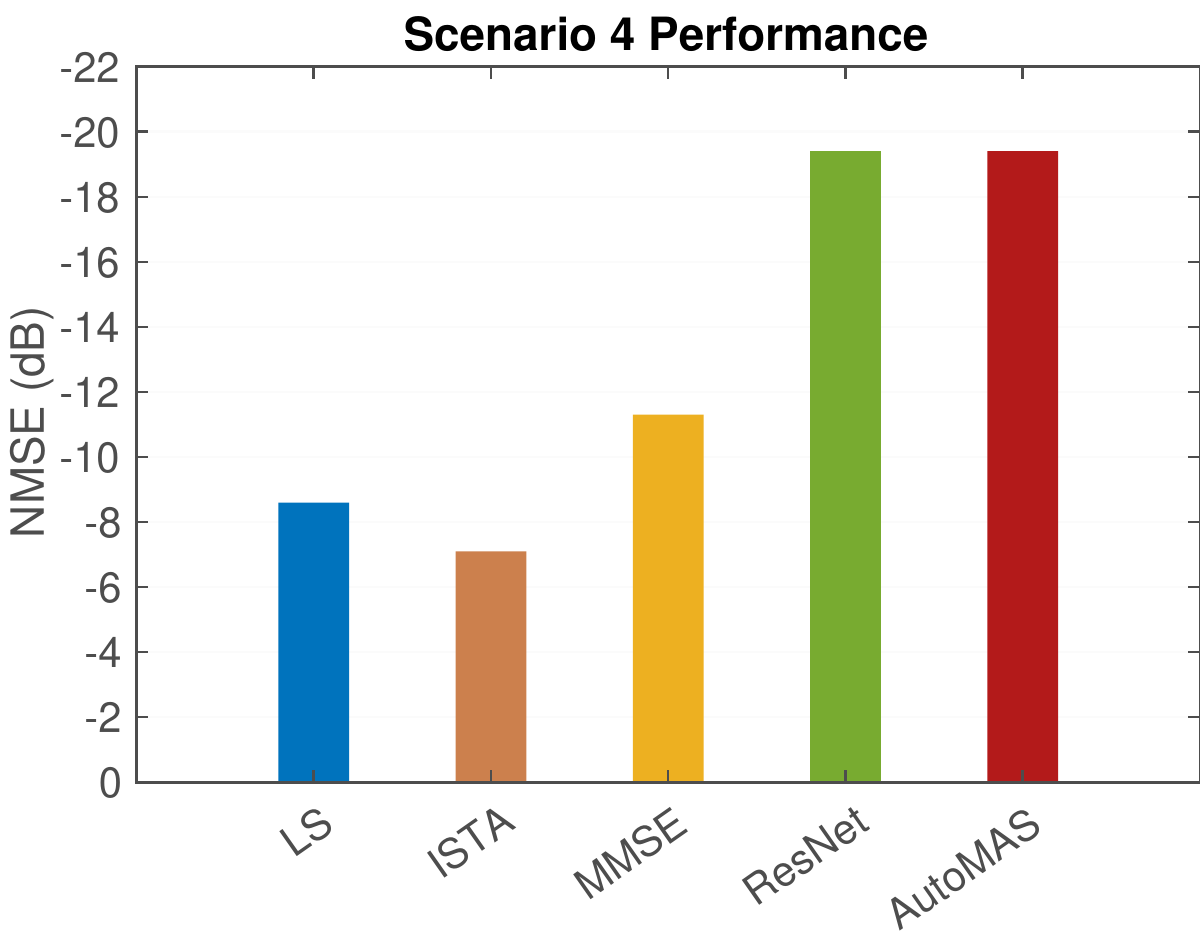}\label{fig:sub4}}

\caption{Comparisons between our AutoMAS and existing channel estimation algorithms (LS, ISTA, LMMSE, ResNet) across different scenarios.}
\label{fig:all}
\end{figure}



%
For the algorithm implementation in the simulation, the ResNet model consists of three residual blocks, each containing two convolutional layers followed by batch normalization. The network is trained using the Adam optimizer with a learning rate of 1e-3 for 150 epochs. For a fair comparison, LMMSE utilizes the covariance matrix of Scenario 3, while ResNet uses the checkpoints of Scenario 4 to test in sample-free scenarios.

As shown in Fig. \ref{fig:all}, fixed algorithm approaches for dynamic channel estimation exhibit many defects. LS performs poorly in the high noise scenario, ISTA struggles when multipath components are not sparse, and LMMSE and ResNet failed in scenarios without access to prior CSI statistics or training data. In contrast, AutoMAS always selects the most suitable algorithm for each scenario. It selects LS, ISTA, LMMSE, and ResNet for Scenarios~1–4, respectively, achieving the best performance across all tested environments.

These findings highlight the dynamic adaptation capability of AutoMAS, enabling seamless handling of increasing heterogeneity and environmental dynamics, which lays a scalable foundation for future wireless network.


\section{Conclusion}
In this work, we propose AutoMAS, a MAS for wireless algorithm self-adaptation through multi-agent collaboration. 
Our system contains a supervisor-executor mechanism for multi-agent coordination, enabling autonomous adjustment to various scenarios. 
A case study on dynamic channel estimation, along with simulations across diverse environments, validates our AutoMAS self-adaptive capabilities, highlighting its generalization for next-generation communication networks.
%
%
Future work will measure and reduce decision-time cost and extend AutoMAS to multi-user and interference-limited settings with joint accuracy–latency optimization.

\bibliographystyle{bibliography/IEEEtran}
\bibliography{bibliography/MAS}

@article{berger2009sparse,
  title={Sparse channel estimation for multicarrier underwater acoustic communication: From subspace methods to compressed sensing},
  author={Berger, Christian R and Zhou, Shengli and Preisig, James C and Willett, Peter},
  journal={IEEE transactions on signal processing},
  volume={58},
  number={3},
  pages={1708--1721},
  year={2009},
  publisher={IEEE}
}

@article{xu2024survey,
  title={A survey on game playing agents and large models: Methods, applications, and challenges},
  author={Xu, Xinrun and Wang, Yuxin and Xu, Chaoyi and Ding, Ziluo and Jiang, Jiechuan and Ding, Zhiming and Karlsson, B{\"o}rje F},
  journal={arXiv preprint arXiv:2403.10249},
  year={2024}
}

@article{wu2023autogen,
  title={Autogen: Enabling next-gen llm applications via multi-agent conversation},
  author={Wu, Qingyun and Bansal, Gagan and Zhang, Jieyu and Wu, Yiran and Li, Beibin and Zhu, Erkang and Jiang, Li and Zhang, Xiaoyun and Zhang, Shaokun and Liu, Jiale and others},
  journal={arXiv preprint arXiv:2308.08155},
  year={2023}
}

@article{guo2024large,
  title={Large language model based multi-agents: A survey of progress and challenges},
  author={Guo, Taicheng and Chen, Xiuying and Wang, Yaqi and Chang, Ruidi and Pei, Shichao and Chawla, Nitesh V and Wiest, Olaf and Zhang, Xiangliang},
  journal={arXiv preprint arXiv:2402.01680},
  year={2024}
}

@article{tong2024wirelessagent,
  title={Wirelessagent: Large language model agents for intelligent wireless networks},
  author={Tong, Jingwen and Shao, Jiawei and Wu, Qiong and Guo, Wei and Li, Zijian and Lin, Zehong and Zhang, Jun},
  journal={arXiv preprint arXiv:2409.07964},
  year={2024}
}

@inproceedings{mongaillard2024large,
  title={Large language models for power scheduling: A user-centric approach},
  author={Mongaillard, Thomas and Lasaulce, Samson and Hicheur, Othman and Zhang, Chao and Bariah, Lina and Varma, Vineeth S and Zou, Hang and Zhao, Qiyang and Debbah, Merouane},
  booktitle={2024 22nd International Symposium on Modeling and Optimization in Mobile, Ad Hoc, and Wireless Networks (WiOpt)},
  pages={321--328},
  year={2024},
  organization={IEEE}
}

@article{zou2023wireless,
  title={Wireless multi-agent generative AI: From connected intelligence to collective intelligence},
  author={Zou, Hang and Zhao, Qiyang and Bariah, Lina and Bennis, Mehdi and Debbah, Merouane},
  journal={arXiv preprint arXiv:2307.02757},
  year={2023}
}

@ARTICLE{quadriga,
  author={Jaeckel, Stephan and Raschkowski, Leszek and Börner, Kai and Thiele, Lars},
  journal={IEEE Transactions on Antennas and Propagation}, 
  title={QuaDRiGa: A 3-D Multi-Cell Channel Model With Time Evolution for Enabling Virtual Field Trials}, 
  year={2014},
  volume={62},
  number={6},
  pages={3242-3256},
  doi={10.1109/TAP.2014.2310220}}

@INPROCEEDINGS{Peaceman,
  author={Yuan, Dingli and Wu, Shitong and Tang, Haoran and Yang, Lu and Peng, Chenghui},
  booktitle={2024 IEEE 24th International Conference on Communication Technology (ICCT)}, 
  title={A Peaceman-Rachford Splitting Approach with Deep Equilibrium Network for Channel Estimation}, 
  year={2024},
  volume={},
  number={},
  pages={1525-1533},
  doi={10.1109/ICCT62411.2024.10946331}}

@inproceedings{kaur2018channel,
  title={Channel estimation in MIMO-OFDM system: a review},
  author={Kaur, Harmandar and Khosla, Mamta and Sarin, RK},
  booktitle={2018 second international conference on electronics, communication and aerospace technology (ICECA)},
  pages={974--980},
  year={2018},
  organization={IEEE}
}

@article{beck2009fast,
  title={A fast iterative shrinkage-thresholding algorithm for linear inverse problems},
  author={Beck, Amir and Teboulle, Marc},
  journal={SIAM journal on imaging sciences},
  volume={2},
  number={1},
  pages={183--202},
  year={2009},
  publisher={SIAM}
}

@INPROCEEDINGS{MMSE,
  author={van de Beek, J.-J. and Edfors, O. and Sandell, M. and Wilson, S.K. and Borjesson, P.O.},
  booktitle={1995 IEEE 45th Vehicular Technology Conference. Countdown to the Wireless Twenty-First Century}, 
  title={On channel estimation in OFDM systems}, 
  year={1995},
  volume={2},
  number={},
  pages={815-819 vol.2},
  doi={10.1109/VETEC.1995.504981}}

@inproceedings{he2016deep,
  author       = {He, Kaiming and Zhang, Xiangyu and Ren, Shaoqing and Sun, Jian},
  title        = {Deep residual learning for image recognition},
  booktitle    = {Proceedings of the IEEE Conference on Computer Vision and Pattern Recognition},
  year         = {2016},
  pages        = {770--778},
}

@article{shao2024deepseekmath,
  title={Deepseekmath: Pushing the limits of mathematical reasoning in open language models},
  author={Shao, Zhihong and Wang, Peiyi and Zhu, Qihao and Xu, Runxin and Song, Junxiao and Bi, Xiao and Zhang, Haowei and Zhang, Mingchuan and Li, YK and Wu, Y and others},
  journal={arXiv preprint arXiv:2402.03300},
  year={2024}
}

@article{ahn2024large,
  title={Large language models for mathematical reasoning: Progresses and challenges},
  author={Ahn, Janice and Verma, Rishu and Lou, Renze and Liu, Di and Zhang, Rui and Yin, Wenpeng},
  journal={arXiv preprint arXiv:2402.00157},
  year={2024}
}

@article{he2018compressive,
  title={Compressive channel estimation and multi-user detection in C-RAN with low-complexity methods},
  author={He, Qi and Quek, Tony QS and Chen, Zhi and Zhang, Qi and Li, Shaoqian},
  journal={IEEE Transactions on Wireless Communications},
  volume={17},
  number={6},
  pages={3931--3944},
  year={2018},
  publisher={IEEE}
}

@article{rost2014cloud,
  title={Cloud technologies for flexible 5G radio access networks},
  author={Rost, Peter and Bernardos, Carlos J and De Domenico, Antonio and Di Girolamo, Marco and Lalam, Massinissa and Maeder, Andreas and Sabella, Dario and W{\"u}bben, Dirk},
  journal={IEEE Communications Magazine},
  volume={52},
  number={5},
  pages={68--76},
  year={2014},
  publisher={IEEE}
}

@article{wang2015artificial,
  title={Artificial intelligence-based techniques for emerging heterogeneous network: State of the arts, opportunities, and challenges},
  author={Wang, Xiaofei and Li, Xiuhua and Leung, Victor CM},
  journal={IEEE Access},
  volume={3},
  pages={1379--1391},
  year={2015},
  publisher={IEEE}
}

@article{jiang2025large,
  title={From large ai models to agentic ai: A tutorial on future intelligent communications},
  author={Jiang, Feibo and Pan, Cunhua and Dong, Li and Wang, Kezhi and Dobre, Octavia A and Debbah, Merouane},
  journal={arXiv preprint arXiv:2505.22311},
  year={2025}
}

\end{document}